# Radio and Optical Spectra of Objects from Two Complete Samples of Radio Sources

## V. Chavushyan[1], R. Mujica[1], A. G. Gorshkov[2], V. K. Konnikova[2], M. G. Mingaliev[3], and J. R. Valdéz[1]

*[1]National Institute of Astrophysics, Optics, and Electronics, Puebla, México*
*[2]Sternberg Astronomical Institute, Universitetskiĭ pr. 13, Moscow, 119899 Russia*
*[3]Special Astrophysical Observatory, Russian Academy of Sciences, Nizhniĭ Arkhyz, 357142 Russia*
Received March 10, 2000

**Abstract**—We present optical identifications and radio spectra for ten radio sources from two flux-density-complete samples. Radio variability characteristics are presented for four objects. The observations were obtained on the RATAN-600 radio telescope at 0.97–21.7 GHz and the 2.1 m telescope of the Haro Observatory in Cananea, Mexico at 4200–9000 Å. Among the ten objects studied, three are quasars, four are BL Lac objects, two are radio galaxies, and one is a Sy 1 galaxy. Two of the sources identified with BL Lac objects, 0509 + 0541 and 0527 + 0331, show rapid variability on time scales of 7–50 days. © 2001 MAIK "Nauka/Interperiodica".

## 1. INTRODUCTION

This paper presents the optical identifications and radio spectra of ten sources from two flux-density-limited complete samples of radio sources, together with variability parameters for four of the sources. The first sample includes all Zelenchuk survey sources with fluxes $S > 200$ mJy at 3.9 GHz, covering 24 h in right ascension, at declinations from $4°$ to $6°$ (B1950) and with Galactic latitudes $|b| > 10°$ [1]. The second sample was derived from the MGB 4.85 GHz survey, and contains sources with flux densities $S > 100$ mJy at declinations between $74°$ and $74°44'$ (J2000).

One reason to study these samples is to try to detect cosmological evolution in the properties of quasars. Currently, it is only for samples with high limiting fluxes, $S > 1$ Jy [2], for which the redshifts of most objects are available. However, such samples contain few distant quasars (the mean redshift in the samples of [2] is $z = 0.7$), making it impossible to investigate evolutionary effects. Our samples have a sufficiently low flux limit to include a significant number of distant quasars. Half of the quasars identified thus far have redshifts $z > 1.4$. In addition, we observe the complete radio luminosity function of quasars up to $z ≈ 1$, making studies of these samples promising for investigations of cosmological relations.

This paper continues our optical identifications for radio sources from the complete samples, initiated in 1998 [3].

## 2. OPTICAL OBSERVATIONS

We obtained the spectra of the program objects in March–August, 1999 using the 2.1 m telescope of the Guillermo Haro Observatory of the National Institute of Astrophysics, Optics, and Electronics in Cananea, Mexico. The observations used an LFOSC spectropho-tometer equipped with a $600 × 400$-pixel CCD [4]. The detector's readout noise was ~8 $e^-$. The wavelength range of the spectrophotometer was from 4200 to 9000 Å, with an 8.2 Å per pixel reciprocal dispersion. The effective instrumental resolution was about 16 Å.

The data reduction was done in the IRAF package and included flat fielding, wavelength linearization, cosmic-ray removal, and flux calibration. The magnitudes for seven objects were taken from the Automated Plate Scanner Catalog of the Palomar Observatory Sky Survey (POSS) [5].

## 3. RADIO OBSERVATIONS

The radio observations were obtained on the southern sector of the RATAN-600 radio telescope with a plane reflector at 3.9 and 7.5 GHz in 1980–1991, and on the northern and southern sectors at 0.97, 2.3, 3.9, 7.7, 11.1, and 21.7 GHz in 1996–1999. The detector parameters and antenna-beam patterns for the northern and southern sectors of RATAN-600 are presented in [6–8]. The sources were observed for 15 to 100 days in each run. The source flux densities were determined by averaging the data from each run. The error in the flux density was determined, as usual, from the scatter of the fluxes measured daily in each observing run. The resulting error includes all relevant errors: noise, calibration error, referencing error of the calibration signal, antenna pointing error, etc. The reduction techniques are described in [9]. The flux density scales for different years were reduced to the scale adopted in [8].

## 4. RADIO AND OPTICAL COORDINATES

Table 1 contains the radio and optical coordinates of the program objects. The source names correspond to hours and minutes of right ascension and degrees and minutes of declination for the equinox J2000. The most accurate coordinates for the studied objects can be found in the Jodrell Bank VLA Astrometric Survey catalogs JVAS2 [10] (8.4 GHz, rms coordinate error 0.014″) and JVAS1 [11] (rms coordinate error 0.012″), and in the NRAO VLA Sky Survey [12] (1.4 GHz, average rms coordinate errors about 0.11″ and 0.56″ in right ascension and declination, respectively). The optical coordinates were derived from the Digitized Palomar Sky Survey.

**Table 1.** Radio and optical coordinates of the program objects

| Object name | Radio coordinates, J2000 | | Optical coordinates, J2000 | | Reference |
|---|---|---|---|---|---|
| 0509 + 0541 | $05^h09^m25^s.97$ | $+05°41'35''.34$ | $05^h09^m25^s.99$ | $+05°41'35''.4$ | JVAS2 |
| 0527 + 0331 | 05 27 32.70 | +03 31 31.50 | 05 27 32.70 | +03 31 31.4 | JVAS2 |
| 0905 + 0537 | 09 05 07.47 | +05 37 16.76 | 09 05 07.46 | +05 37 16.6 | NVSS |
| 1027 + 7440 | 10 27 39.10 | +74 40 04.7 | 10 27 39.20 | +74 40 04.4 | NVSS |
| 1243 + 7442 | 12 43 45.03 | +74 42 37.13 | 12 43 44.90 | +74 42 38.0 | JVAS1 |
| 1411 + 7424 | 14 11 34.74 | +74 24 29.1 | 14 11 34.75 | +74 24 29.1 | NVSS |
| 1424 + 0434 | 14 24 09.50 | +04 34 52.06 | 14 24 09.58 | +04 34 51.3 | JVAS2 |
| 1426 + 0426 | 14 26 28.92 | +05 26 58.12 | 14 26 28.99 | +04 26 58.2 | NVSS |
| 1511 + 0518 | 15 11 41.27 | +05 18 09.26 | 15 11 41.28 | +05 18 09.1 | JVAS2 |
| 1923 + 7404 | 19 23 23.04 | +74 04 04.9 | 19 23 23.48 | +74 04 05.1 | NVSS |

## 5. RESULTS

Figures 1 and 2 present the radio and optical spectra of the objects. Below, we quote the old name of the source, corresponding to B1950 coordinates, in parentheses.

### *0509 + 0541 (0506 + 056)*

We observed this source with the RATAN-600 telescope in eight runs at 3.9 and 7.5 GHz in 1980–1991. The highest flux density, $S_{(3.9)} = 894 \pm 29$ mJy and $S_{(7.5)} = 895 \pm 65$ mJy, was recorded in October 1984, and the lowest flux densities, in August 1991, were $S_{(3.9)} = 450 \pm 25$ mJy and $S_{(7.5)} = 420 \pm 22$ mJy (rms errors are quoted here). The weighted average flux densities, $\langle S \rangle$, during our observations were 536 and 531 mJy, respectively, at 3.9 and 7.5 GHz. The variability index for 1984–1991, $V = dS/\langle S \rangle$, derived taking into account individual errors as described in [1, 13], was 0.5 at both frequencies.

Starting in 1996, we observed the source at six frequencies. In 1999, we included it in our program to search for flux density variability on short time scales, and it was observed daily for 100 days starting on May 22, 1999. Figure 1 shows the radio spectra of the source for August 1997 and July 1999. The substantial variability at low frequencies (0.97 and 2.3 GHz) is striking. This source has a small extended component, while its compact component makes the main contribution to its radiation. In the August 1997 spectrum, the peak flux density is at about 12 GHz. The source was in a more active state in 1999; its peak flux density shifted toward higher frequencies and the spectrum could be approximated with a logarithmic parabola, $\log S = 2.727 + 0.173 \log \nu - 0.044 \log^2 \nu$, where the flux density is in mJy and the frequency in GHz. During this interval, the source showed variability at 3.9 and 2.3 GHz on time scales less than ten days, as well as cyclic variability with a 52 day quasi-period at the same frequencies (and possibly also at 7.7 GHz). The cyclic variability at the different frequencies is correlated.

The source was not resolved with the VLA B [14]. In [14], it was identified with a star-like object with magnitudes $16^m$ and $15.5^m$ on the POSS *O* and *E* prints. In 1992, at our request, the object's spectrum was taken with the 6-m telescope of the Special Astrophysical Observatory [15], and it was tentatively identified as a BL Lac object. The optical spectrum in Fig. 1 obtained on March 15, 1999 with an exposure time of 40 min shows a featureless continuum, supporting this identification.

### *0527 + 0331 (0524 + 034)*

This is a unique source, with the largest amplitude for long-term flux-density variability among all known radio sources. Its flux density at 7.7 GHz increased more than 20-fold from 1988 till 1998. The pattern of its long-term variability is described in [16]; the variability index for 1984–1991 was $V = dS/\langle S \rangle = 1.9$. In addition, appreciable variability on time scales less than ten days was detected in observations from January 3 to February 25, 1998 [17]. During those observations, the source's activity state was nearly at maximum; its mean spectrum for that period is shown by the crosses in Fig. 1. The flux density increased towards higher frequencies.

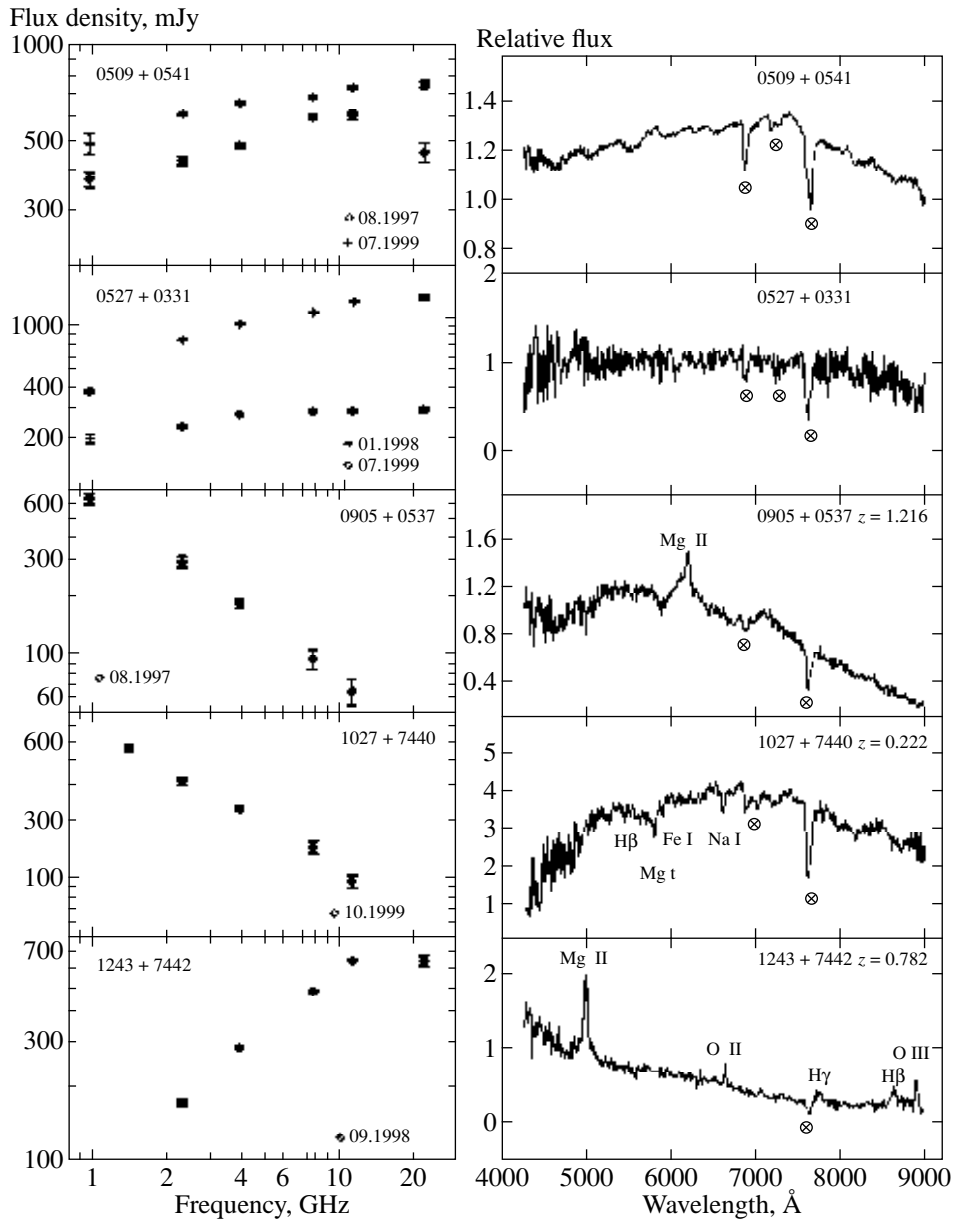

**Fig. 1.** Radio and optical spectra of 0509 + 0541, 0527 + 0331, 0905 + 0537, 1027 + 7440, and 1243 + 7442.

Starting on May 22, 1999, we observed the source daily over 100 days; the mean spectrum for this time interval is indicated by the diamonds in Fig. 1. The source's flux at all frequencies dropped almost five-fold, but the relative amplitude of the rapid variability and the time scales remained unchanged.

The source was identified with a $20^m$ object on the POSS $O$ plate. An optical spectrum taken on October 13,

1999 with a 60-min exposure shows a featureless continuum, suggesting that it is a BL Lac object.

### 0905 + 0537 (0902 + 058)

This source has a constant flux density, and a power-law spectrum from 0.97 to 11.1 GHz: $S = 627v^{-0.920}$ mJy. It was identified with a star-like object with magnitudes $18.1^m$ and $16.9^m$ on the POSS $O$ and $E$ prints. An optical spectrum with a 60-min exposure obtained on March 15, 1999 showed a single strong emission line at 6200 Å, which was interpreted as the Mg II 2798 Å line at a red-shift of $z = 1.216$. Thus, the object can be identified as a quasar.

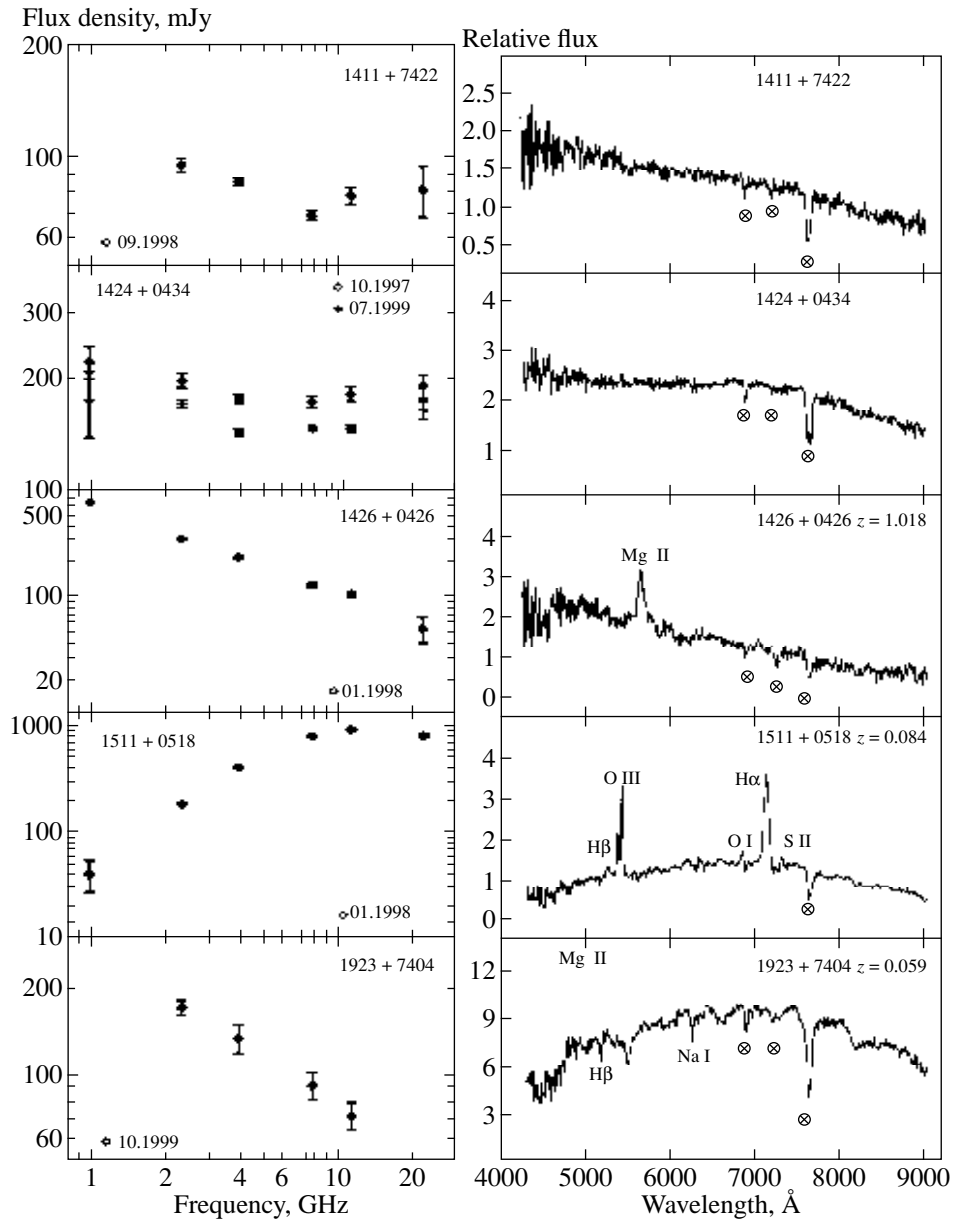

**Fig. 2.** Radio and optical spectra of 1411 + 7422, 1424 + 0434, 1426 + 0426, 1511 + 0518, and 1923 + 7404.

### 1027 + 7440

This source has a power-law spectrum from 0.97 to 11.1 GHz, $S = 597\nu^{-0.728}$ mJy. The corresponding optical object has magnitudes $17.8^m$ and $14.1^m$ on the POSS *O* and *E* prints. An optical spectrum taken on June 3, 1999 with an exposure time of 35 min revealed four absorption lines corresponding to Hβ 4861 Å, Mg I 5175 Å, Fe I 5270 Å, and Na I 5986 Å at a redshift of $z = 0.122$. This spectrum suggests this source should be classified as a radio galaxy.

### 1243 + 7442

The radio spectrum of this source was observed in September, 1998. It can be approximated by a logarithmic parabola at 2–21 GHz, $\log S = 1.649 + 1.729 \log\nu - 0.608 \log^2\nu$. The peak flux density is at about 15 GHz.

The radio source was identified with a star-like object with magnitudes $19.3^m$ and $18.6^m$ on the POSS *O* and *E* prints. An optical spectrum taken on June 4, 1999 with a 60-min exposure shows emission lines at 4986, 6641, 7734, 8662, and 8923 Å, which correspond to Mg II 2798 Å, [O II] 3727 Å, Hγ 4340 Å, Hβ 4861 Å, and [O III] 5007 Å lines at a redshift of $z = 0.782$. We therefore classify the object as a quasar.

### 1411 + 7421

Our radio spectrum obtained in September 1998 has a flux-density minimum at about 4 GHz, and can be approximated with a logarithmic parabola from 2.3 to 21.7 GHz, $\log S = 2.188 - 0.651 \log \nu + 0.321 \log^2 \nu$. Two components are present in the spectrum: an extended one with a power-law spectrum and a compact one with a spectrum that grows toward higher frequencies.

The optical spectrum of 1411 + 7421 was taken on June 3, 1999, with an exposure time of 40 minutes. The POSS $O$ and $E$ magnitudes of the object are $17.2^m$ and $16.6^m$. Since we detected no lines in the spectrum, we classify this source as a BL Lac object.

### 1424 + 0434 (1421 + 048)

This source has been observed at 3.9 and 7.5 GHz since 1980. During this time, its flux slowly decreased at both frequencies (at 3.9 GHz, from 330 mJy in 1980 to 143 mJy in 1999). The flux variability index for 1984–1991, $V = dS/\langle S \rangle$, is 0.25 at both frequencies.

Since 1996, we have observed the source at six frequencies. Figure 2 presents its spectra for November 1997 and July 1999. Both spectra show a minimum at 5–6 GHz, and the flux decrease occurred without significant changes in the appearance of the spectrum. The July 1999 spectrum corresponds to a logarithmic parabola $\log S = 2.341 - 0.437 \log \nu + 0.259 \log^2 \nu$. Our daily observations covering 100 days starting on May 22, 1999 did not reveal any significant flux density variability on time scales less than several weeks.

The radio source was identified with a star-like object with magnitudes $20.1^m$ and $18.6^m$ on the POSS $O$ and $E$ prints. The object's optical spectrum obtained on June 4, 1999 with an exposure time of 40 min did not reveal any detectable lines, leading us to classify the source as a BL Lac object.

### 1426 + 0426 (1423 + 046)

This source has a constant flux density and a power-law spectrum from 0.365 to 21.7 GHz, $S = 610\nu^{-0.767}$ mJy. VLA observations at 4.85 GHz show the radio source to be triple [14], and it is identified with a star-like optical object coinciding with the radio center. The object's magnitudes on the POSS $O$ and $E$ prints are $19.1^m$ and $18.1^m$. A single emission line is present in a spectrum taken on June 5, 1999 with a 40-min exposure, corresponding to the Mg II 2789 Å line at a redshift of $z = 1.018$. We accordingly classify the object as a quasar.

### 1511 + 0518 (1509 + 054)

Figure 2 shows a spectrum from 0.97 to 21.7 GHz taken in January 1998. It is well approximated with a logarithmic parabola, $\log S = 1.410 + 2.756 \log \nu - 1.222 \log^2 \nu$, with a peak flux density at about 13 GHz. Unusual features of the spectrum include the absence of an extended component and the large spectral index at frequencies where the spectrum is growing (at 0.97–3.9 GHz, $\alpha = 1.58$, $S \propto \nu^\alpha$). We observed the source at 3.9 and 7.5 GHz in 1984–1990, during which time we detected no statistically significant variability at these frequencies. Our subsequent observations at six frequencies show modest flux variability at frequencies above 7.7 GHz; the variability index for 1.5 years (1997–1998) at 11.1 GHz is $V = 0.1$. A 60-day search for rapid variability in 1998 did not show any appreciable changes of the source's radio flux on time scales less than several weeks.

The corresponding optical object has magnitudes $17.7^m$ and $15.3^m$ on the POSS $O$ and $E$ prints. The optical spectrum in Fig. 2 was taken on August 7, 1999 with a 20-min exposure. The observed system of lines corresponds to Hβ 4861 Å, forbidden nebular [O III] 4959 and 5007 Å, forbidden [O I] 6300 Å, the Hα 6563 Å emission lines at a redshift of $z = 0.084$. In addition, the forbidden [S II] 6717/6731 Å doublet is present, but the spectral resolution was insufficient to separate its components. The width of the hydrogen lines is FWHM $\approx$ 3000 km/s; while that of the forbidden lines is FWHM $\approx$ 1000 km/s. We classify the object as a Seyfert galaxy of type SyI at a redshift of $z = 0.084$.

### 1923 + 7404

This object has a power-law spectrum from 2.3 to 11.1 GHz, with spectral index $\alpha = -0.58$. The flux density at 3.9 GHz is $133 \pm 10$ mJy.

In the optical spectrum of 1923 + 7404 obtained on June 4, 1999 with an exposure time of 30 min, three absorption lines are apparent: Hβ 4861 Å, Mg I 5175 Å, and Na I 5896 Å, at a redshift of $z = 0.059$. We therefore classify the object as a radio galaxy; it is identified with a $16^m$ object on the POSS $O$ print.

## 6. CONCLUSIONS

Of the ten objects studied, three proved to be quasars, four BL Lac objects, two radio galaxies, and one a Seyfert galaxy. Two of the BL Lac objects, 0509 + 0541 and 0527 + 0331, show rapid variability on time scales of 7–50 days.

Table 2 summarizes some of the results of our optical and radio observations. Its columns contain (1) the source names, (2) the lines present in the optical spectra, (3) the rest and observed wavelengths of the lines, (4) the redshifts, (5) the object classifications, (6) the POSS $O$ magnitudes, (7) the radio flux densities at 3.9 and 11.1 GHz, and (8) the spectral index $\alpha$ between these frequencies.

**Table 2.** Summary of optical and radio observations

| Object name | Spectral Lines | Wavelength, Å | $z$ | Classification from the spectrum | $m_O$ | Flux density at 3.9 and 11.1 GHz in 1998, mJy | $\alpha$ |
|---|---|---|---|---|---|---|---|
| 1 | 2 | 3 | 4 | 5 | 6 | 7 | 8 |
| 0509 + 0541 | None | | | BL Lac | 16.0 | $657 \pm 7$ | 0.11 |
| | | | | | | $734 \pm 9$ | |
| 0527 + 0331 | None | | | BL Lac | 20.0 | $886 \pm 7$ | 0.31 |
| | | | | | | $1228 \pm 15$ | |
| 0905 + 0537 | Mg II | 2798/6200 | 1.216 | QSO | 18.1 | $180 \pm 10$ | −0.97 |
| | | | | | | $65 \pm 10$ | |
| 1027 + 7440 | Hβ | 4861/5454 | 0.122 | Gal | 17.8 | $225 \pm 7$ | −0.81 |
| | Mg I | 5175/5806 | | | | $96 \pm 7$ | |
| | Fe I | 5270/5913 | | | | | |
| | Na I | 5896/6615 | | | | | |
| 1243 + 7442 | Mg II | 2798/4986 | 0.782 | QSO | 19.3 | $284 \pm 3$ | 0.80 |
| | [O II] | 3727/6641 | | | | $654 \pm 6$ | |
| | Hγ | 4340/7734 | | | | | |
| | Hβ | 4861/8662 | | | | | |
| | [O III] | 5007/8923 | | | | | |
| 1411 + 7424 | None | | | BL Lac | 17.2 | $85 \pm 2$ | −0.08 |
| | | | | | | $78 \pm 4$ | |
| 1424 + 0434 | None | | | BL Lac | 20.1 | $176 \pm 5$ | 0.03 |
| | | | | | | $181 \pm 8$ | |
| 1426 + 0426 | Mg II | 2798/5646 | 1.018 | QSO | 19.1 | $215 \pm 6$ | −0.70 |
| | | | | | | $103 \pm 4$ | |
| 1511 + 0518 | Hβ | 4861/5269 | 0.084 | Sy1 | 17.7 | $407 \pm 5$ | 0.77 |
| | [O III] | 4959/5376 | | | | $906 \pm 13$ | |
| | [O III] | 5007/5428 | | | | | |
| | [O I] | 6300/6829 | | | | | |
| | Hα | 6563/7114 | | | | | |
| | [S II] | 6724/7289 | | | | | |
| 1923 + 7404 | Hβ | 4861/5148 | 0.059 | Gal | 16.0 | $133 \pm 15$ | −0.58 |
| | Mg I | 5175/5480 | | | | $72 \pm 8$ | |
| | Na I | 5896/6244 | | | | | |

## ACKNOWLEDGMENTS

The authors are very grateful to the administration of the G. Haro Observatory (Mexico) for their support of and attention to this study. This work was supported by the Russian Foundation for Basic Research (project code 98-02-16428), the "Universities of Russia" program (grant 5561), the State Scientific and Technology Program "Astronomy" (project 1.2.5.1), and, in part, CONACYT (grants 28499-E and J32178-E).

*Translated by N. Samus'*